\documentclass[a4paper, oneside, twocolumn, notitlepage, 10pt]{extarticle_ecoc}
\usepackage{ecoc}
\usepackage{comment}
\usepackage{graphicx}
\begin{document}
\selectlanguage{english}    


\title{LLMs and Optical Networks: A Symbiotic Relationship}


\author{
    Mëmëdhe Ibrahimi, Qiaolun Zhang, Giovanni S. Sticca, Jiaheng Xiong, \\Francesco Musumeci, and Massimo Tornatore
}

\maketitle                  


\begin{strip}
    \begin{author_descr}

        Politecnico di Milano,
        \textcolor{blue}{\uline{massimo.tornatore@polimi.it}} 

    \end{author_descr}
\end{strip}

\renewcommand\footnotemark{}
\renewcommand\footnoterule{}

\begin{figure*}[b]
    \centering
    \includegraphics[width=\textwidth]{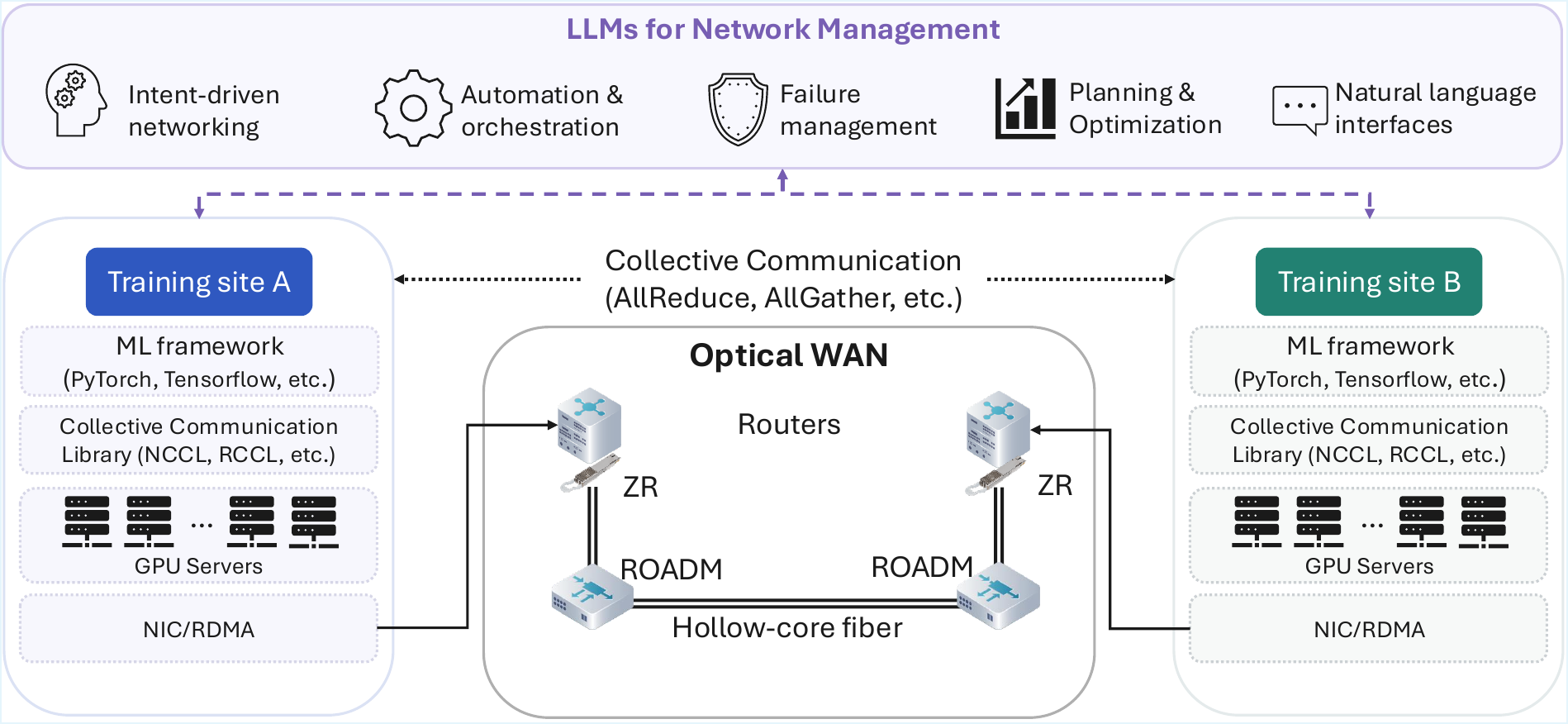}
    \caption{LLMs and optical networks: a system overview.}
    \vspace{-30pt}
    \label{fig:overview}
\end{figure*}
\begin{strip}
    \begin{ecoc_abstract}
    This paper explores the emerging symbiosis between LLMs and optical networks. Massive LLMs require geo-distributed training, which demands advanced optical transport capabilities that require new key technical enablers, as WAN-aware CCL algorithms, ZR+ pluggables, and Hollow Core Fibers.  Conversely, LLMs also enable new forms of autonomous network management. \textcopyright 2026 The Author(s)
    \end{ecoc_abstract}
\end{strip}

\section{Introduction: The Need for Geo-Distributed AI}
The scale of Large Language Models (LLMs) is increasing at an unprecedented pace, and so is the infrastructure required for training and inference. The most advanced AI models are expected to require training platforms comprising on the order of $10^5$ GPUs, pushing system design beyond the practical limits of a single data center (DC)~\cite{dong2025beyond}. 

Co-locating compute resources of this magnitude at a single DC raises severe constraints, including: i) \emph{power density}, as a 100K-GPU cluster demands 100--150 MW, easily exceeding local grid capacities; ii) \emph{cost and resource availability}, bottlenecked by multi-billion-dollar investments, GPU shortages, and the scarcity of sustainable sites; and iii) \emph{data locality and regulations}, which often prohibit centralizing globally distributed datasets. 

Naturally, AI infrastructures rely on \emph{scale-out} (or horizontal scaling), an architectural approach that increases system capacity by adding more computing nodes, such as servers or GPUs, to work in parallel within a single DC. However, the aforementioned limitations are driving a mandatory industry transition toward \emph{scale-across}, an AI networking architecture that connects geographically distributed DCs into a unified ``AI super factory''~\cite{allison2025scaleacross}.

In this distributed paradigm, the primary challenge shifts from computation to moving and synchronizing massive data volumes across Wide Area Networks (WANs). Unfortunately, distributed training is inherently straggler-sensitive, as a training iteration's completion time is dictated by the slowest inter-site link. Hence, latency, heterogeneous paths, and limited control degrade end-to-end training efficiency. This challenge is especially severe outside hyperscaler backbones (e.g., in public or multi-domain WANs), where routing, capacity, and optical configurations are only partially controllable~\cite{xu2026impairment-sharing}. 

To overcome these bottlenecks, a fundamental evolution of the optical networking ecosystem is required. Geo-distributed AI training must be addressed as a joint communication and coordination problem that spans from the logical layer (i.e., algorithms) down to the physical transmission media. At the logical level, Collective Communication Libraries (CCLs) must become WAN-aware, intelligently scheduling tensor transfers across asymmetric optical links. At the architectural and physical layers, the network must scale to support unprecedented east-west traffic volumes. This involves deploying energy-efficient coherent optics, such as ZR/ZR+ pluggables, to seamlessly interconnect regional data center clusters, as well as adopting paradigm-shifting transmission media, such as Hollow Core Fibers (HCF), to achieve ultra-low latency and ultra-high capacity.

However, the relationship between optical networks and LLMs is not unidirectional, but rather it is symbiotic. While optical networks provide the critical ``\emph{Networking for AI}'' infrastructure required to train foundational models, LLMs provide the ``\emph{AI for Networking}'' intelligence to manage complex physical topologies. LLMs are emerging as powerful agents capable of interpreting operator intents, automating complex configurations, and performing root-cause analysis for network failures. 

Fig.~\ref{fig:overview} illustrates this symbiotic setting. Training sites exchange massive volumes of traffic over WANs, making optical networking the enabler of scalable, geo-distributed AI. Simultaneously, LLM-powered agents sit atop the control plane, streamlining the orchestration and resilience of the transport infrastructure. 
In this paper, we explore this symbiotic relationship. First, we detail the critical optical enablers, including WAN-aware CCL algorithms, ZR+ pluggables, and HCF, that support geo-distributed AI training. We then review the reverse symbiosis, illustrating how LLMs enable autonomous network management.

\section{Algorithmic Innovations: WAN-Aware Collective Communication}
For distributed LLM training, network communication has emerged as a first-order bottleneck within the \emph{scale-out} boundaries of a single DC, where each iteration moves large tensors across GPUs~\cite{liu2024rethinkingmachine,anon2025efficientdirect,cao2025sycclexploiting}. Once training is stretched across regional DCs into a \emph{scale-across} architecture, as in recent geo-distributed and optical-transport training ~\cite{li2024acceleratinggeo,dai2025geopipegeo}, this bottleneck becomes even more stringent. Synchronous collective steps must wait for the slowest participating transfer, so a single congested inter-regional segment can idle otherwise fast GPU groups~\cite{xu2026multipath}. At the algorithmic layer, these exchanges are expressed as \emph{collective communication} primitives, such as AllReduce, AllGather,etc. 
A CCL, such as NCCL ~\cite{anon2017nvidiacollective}, is the runtime that implements these primitives by selecting algorithms, partitioning tensors, scheduling transfers, and mapping traffic onto the network~\cite{liu2024rethinkingmachine}.
Recent CCL synthesizers improved schedule quality in tightly coupled GPU fabrics~\cite{anon2025efficientdirect,cao2025sycclexploiting}. 
However, they were mainly optimized for uniform, high-bandwidth, and low-latency intra-DC fabrics, where bandwidth is largely stable. Their default CCL schedules can therefore underutilize WAN capacity and amplify stragglers~\cite{xiong2025scaleccl, xiong2026coordinatingccl}.


Looking beyond the scope of private WANs managed by single hyperscalers with centralized, end-to-end control~\cite{,jain2013b4}, our target is the broader environment of public WAN connectivity across regional DCs.
A geographically-distributed inter-DC operation relying on public WAN finds an application in multiple contexts, e.g., 
universities, research consortia, national AI infrastructures, smaller cloud providers, and enterprises needing to pool distributed GPUs without owning a global backbone. It is also vital when data sovereignty, energy, GPU supply, or collaboration constraints prevent moving compute into a single operator-controlled fabric. In these cases, traffic crosses leased circuits, carrier networks, or multi-provider interconnects whose internal routing and optical-layer decisions remain outside the operator's direct control~\cite{feamster2014road}.

\begin{figure}[htpb]
    \centering
    \includegraphics[width=\linewidth]{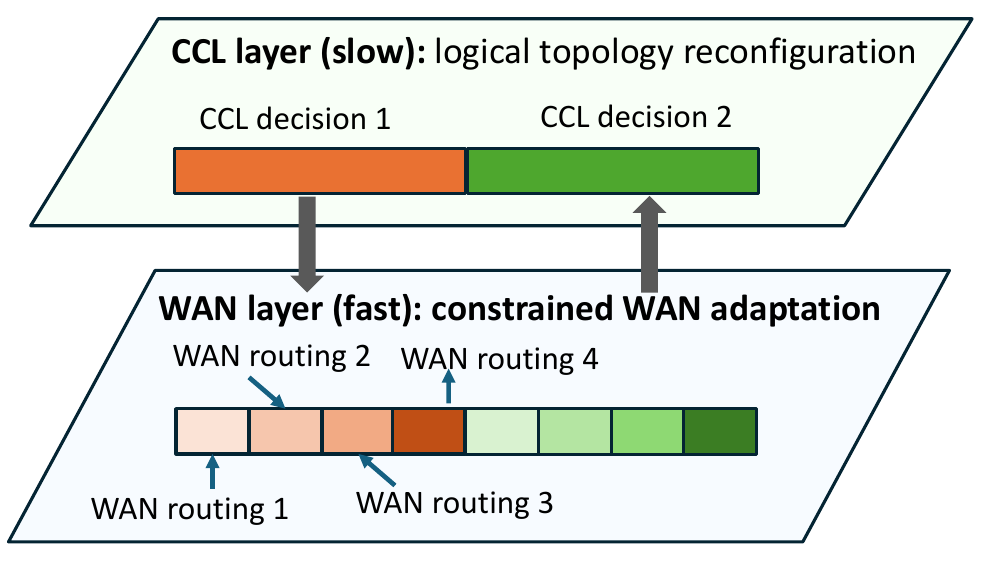}
    \caption{Logical topology reconfiguration and WAN adaptation.}
\label{fig:slow_fast}
\end{figure}



This incomplete control leads to imperfect network observability and restricts the training system to a limited set of routing and service knobs. WAN-aware training should therefore treat the network as a partially-observable and partially-controllable substrate~\cite{xiong2026coordinatingccl}. 
The bottleneck is thus both a bandwidth problem and a coordination problem. The CCL layer adapts slowly by rebuilding logical topology and schedules to mitigate persistent stragglers~\cite{liu2024rethinkingmachine,xiong2025scaleccl}, while the WAN layer reacts faster through limited mechanisms such as overlays, tunnels, and service classes~\cite{hong2013achievinghigh}. These two layers must be coordinated without assuming private-WAN-style end-to-end control. 
Fig.~\ref{fig:slow_fast} illustrates a two-timescale cross-layer control framework, as proposed \cite{xiong2026coordinatingccl}. The CCL layer operates more slowly and performs occasional logical-topology reconfiguration, whereas the WAN layer reacts more frequently through constrained routing adaptation. Key message is that WAN-aware training depends on coordinating slow collective-layer restructuring with fast but limited WAN-layer reroutings.

\section{Optical Network Enablers for the AI-WAN}
This section focuses on two critical physical-layer technologies, ZR/ZR+ pluggables and HCF, that can effectively enable  massive data exchanges of geo-distributed LLM training,.

\textbf{\emph{ZR/ZR+ Pluggables}}. 
Integrating coherent optical transponders directly into IP routers (IPoWDM) via ZR/ZR+ pluggables yields in space, cost, and power consumption reduction by up to 35\% compared to Long-Haul transponders~\cite{ZR-Qiaolun}. Their deployment is synergetic with the modern evolution of AI infrastructure through several key trends: 1) \emph{Enabling Multi-DC Regional Networks}. Driven by power and land scarcity, cloud providers are replacing monolithic ``mega-data centers'' with ``regions'' (clusters) of smaller DCs spaced tens of kilometers apart~\cite{Meta-ECOC25}. This distributed topology lays the groundwork for \emph{scale-across} networking and aligns with the short-reach capabilities of ZR/ZR+ pluggables;  
2) \emph{Capacity Scaling}. The rapid evolution of ZR modules, scaling from 400G up to 1.6T data rates, ensures they can absorb the massive, exponential growth of intra-region traffic; and 3) \emph{Multi-Vendor Interoperability}. 
Unlike proprietary high-performance transponders that require the same manufacturer's equipment at both ends of a link, ZR pluggables from different vendors can interoperate seamlessly. 

A similar (and even tighter) integration step is bing adopted within the DC with the  \emph{Co-Packaged Optics} (CPO), which mounts transceivers directly on the switch package to reduce electrical losses, unlocking higher port density and lower power-per-bit. However, despite their cost and energy efficiencies, both CPO and ZR pluggables are constrained by limited transmission reach. 
To extend their utility (e.g.m in case of ZR, beyond regional deployments), networks deployments are expected to combine these optics along HCF, which we discuss next.

\textbf{\emph{Hollow Core Fibers (HCF)}} overcome the fundamental limits of Standard Single-Mode Fibers (SSMF) by transmitting optical signals through an air-filled core. Compared to SSMF, HCF introduces three main advantages: 1) \emph{lower latency} by $\sim$30\% as light propagates faster in air; 2) \emph{lower transmission loss}, reaching 0.11--0.05 dB/km~\cite{Chen_24, YOFC-004}; and 3) \emph{lower nonlinearity} by 3--4 orders of magnitude, enabling high power amplification. Such benefits provide network-level gains across several use-cases: 1) \emph{Edge Data Center Consolidation}. The strict latency constraints of emerging 6G and distributed AI applications are forcing the deployment of numerous, distributed edgeDCs. By leveraging HCF's latency reduction, network operators can meet these requirements while physically consolidating their infrastructure (e.g., upgrading roughly 20\%--30\% of a network's links to HCF can reduce the number of edgeDCs by up to 30\%~\cite{OFC24,TNSM-polimi}); 2) \emph{Capacity Scaling}. The wide usable bandwidth and low transmission loss of HCF make it a plausible solution for long-term capacity scaling. Deploying HCF as a hybrid or full replacement for multi-fiber bundles can yield significantly higher traffic (e.g., delivering over 2x the throughput of traditional parallel SSMF or multiband networks~~\cite{polimi-OFC25}); and 3) \emph{High-Power Amplification}. As HCF can support high optical launch powers without nonlinear penalties, networks can use high-power EDFAs to increase network throughput and reduce the number of inline amplification sites, yielding reductions of up to 50\% in power consumption per Tbps~\cite{10926228}.
\begin{figure}
    \centering
    \includegraphics[width=1.0\linewidth]{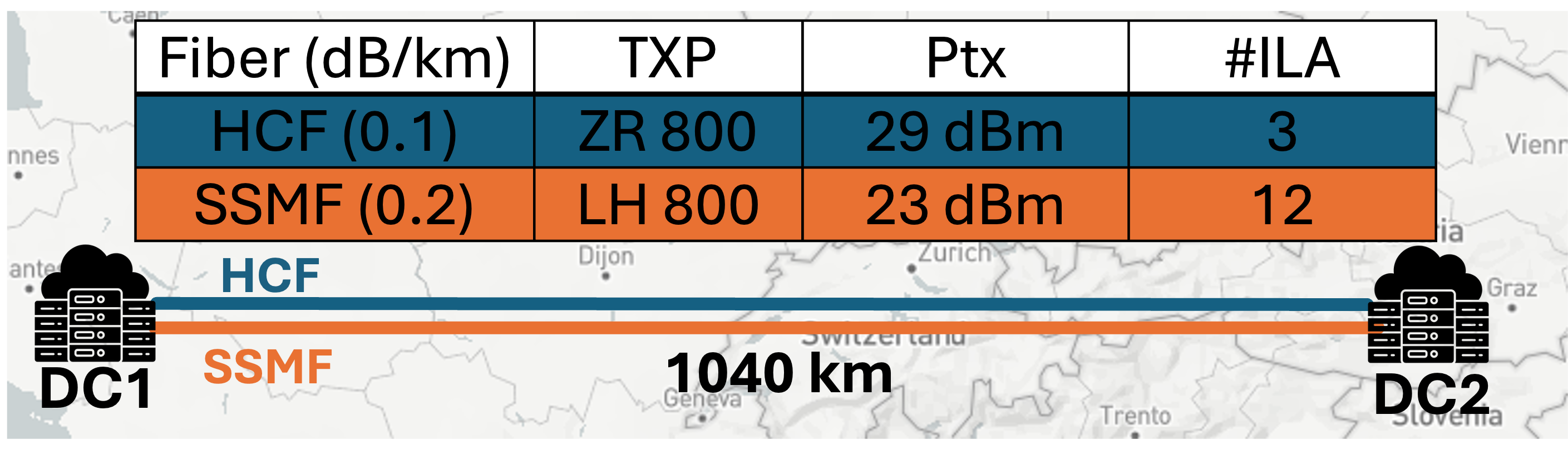}
    \caption{Inter-DC connection with SSMF and HCF}
    \label{fig:ZRandHCF}
\end{figure}
Moreover, HCF is an enabler to extend the inherent reach limitations of ZR/ZR+ pluggables. Fig.~\ref{fig:ZRandHCF} illustrates an example of the synergy between ZR and HCF, in a long-distance inter-DC connection for two scenarios: 1) with SSMF, requiring the deployment of power-hungry LH transponders and 12 amplification sites (e.g., every 80 km) with maximum amplification output power of 23 dBm, and 2) with HCF, enabling the deployment of energy-efficient ZR pluggables and resulting in 3 amplification sites (every 260 km) with maximum amplification power of 29 dBm.  

Beyond HCF and ZR pluggables, latency-sensitive AI workloads can further benefit from \emph{Optical Circuit Switching} (OCS), bypassing packet-layer queuing and establishing dedicated lightpaths with deterministic, microsecond-level propagation delays. However, a thorough treatment of OCS is beyond the scope of this paper.

\section{The Reverse Symbiosis: LLMs for Network Management}
A ``reverse symbiosis'' is rapidly emerging: using LLMs to automate the very optical networks that support them. The industry is shifting from rigid rules to intent-driven autonomy, with LLM agents revolutionizing three primary domains of network management: 1) \emph{Network Configuration and Intent Translation}. LLMs simplify complex interfaces (e.g., NetConf, TAPI) by enabling Intent-Based Networking, translating natural-language intents into vendor-specific configurations~\cite{10926059, 11263212}. 
Additionally, Retrieval-Augmented Generation (RAG) enables data-sovereign automation in multi-vendor Open Optical Transport Networks without extensive model fine-tuning~\cite{11140539, 11046468}; 
2) \emph{Lifecycle Management and Power Optimization}. To ensure data privacy and low latency, recent demonstrations deploy local, lightweight AI agents to manage the full network lifecycle~\cite{10977747}. 
By combining human expertise with digital twins (DT) for safe evaluation, LLMs that act as reasoning agents autonomously optimize optical launch powers to achieve target Quality of Transmission (QoT) in field-deployed networks~\cite{11494190, 11481489}; and 3) \emph{Fault Diagnosis and Multi-Agent Orchestration}. To manage high-volume alarms during physical failures, hybrid frameworks combine graph-based algorithms with LLMs to enable root-cause localization. Using graph-derived Chain-of-Thought (CoT) reasoning, LLMs can diagnose complex failure scenarios, overcoming traditional ``black box'' limitations~\cite{11012654}. 

\clearpage

\defbibnote{myprenote}{%
}
\printbibliography[prenote=myprenote]

\vspace{-4mm}

\end{document}